# THE HEALTH STATUS OF A POPULATION ESTIMATED—THE HISTORY OF HEALTH STATE CURVES

Christos H Skiadas[1] and Charilaos Skiadas[2]

[1]ManLab, Technical University of Crete, Chania, Crete, Greece (Skiadas@cmsim.net )
[2]Department of Mathematics and Computer Science, Hanover College, IN, USA (Skiadas@hanover.edu )

**Preliminaries**

Following the recent publication of our book on "Exploring the Health State of a Population by Dynamic Modeling Methods" (DOI 10.1007/978-3-319-65142-2) in "The Springer Series on Demographic Methods and Population Analysis" we provide this brief presentation of the main findings and improvements regarding the Health State of a Population. (See at: http://www.springer.com/gp/book/9783319651415 ).

Here the brief history of the Health State or Health Status curves for individuals and populations is presented including the main references and important figures along with an illustrated Poster (see Figure 13).

Although the Survival Curve is known as long as the life tables have introduced, the Health State Curve was calculated after the introduction of the advanced stochastic theory of the first exit time.

The health state curve is illustrated in several graphs either as a fit curve to data or produced after a large number of stochastic realizations. The Health State, the Life Expectancy and the age at "mean zero health state" are also estimated.

**The main parts of the stochastic theory needed**

The Health State of an individual is a stochastic process by means that is highly unpredictable in the time course while fluctuates from higher to lower values. However, we have detailed and accurate data sets for deaths over time, that is we know the distribution of people reaching the end (the death distribution of a population) when the health state of an individual is reaching the zero level.

In a modeling approach the problem can be set as a first exit or hitting time modeling of a stochastic process crossing a barrier. Technically the stochastic theory was developed during last two centuries with the observation of the so-called Brownian motion and proved experimentally by Jean Perrin while the mathematical modeling of this classical stochastic process was due to several scientists including Thorvald N. Thiele, James Clerk Maxwell, Stefan Boltzmann, Albert Einstein, Marian Smoluchowski, Paul Lévy, Louis Bachelier. Mathematically the simple stochastic process of the Brownian motion is presented as a Wiener stochastic process in honor of Norbert Wiener. In nowadays the Wiener process is included in the majority of computing devices and programs thus giving the opportunity to generate stochastic paths and simulate processes as the health state of an individual.

So far after almost 150 years of quantitative works on stochastic theory the problem remains unsolved in the microscopic level. It is not possible to know exactly the development of the health state of an individual in the course of time as it is not possible to know the place and speed of a particle in a gas.

Fortunately that we have learned is that we can find "mean properties" of "large ensamples". We can estimate the pressure of a gas in a box, the temperature, the "mean speed" of the molecules and so-on. Accordingly we can find the "health state" or the "operational state" or the "viability" of a population as the "mean health state" of the total number of individuals. That we know is the effect on the population when the health state of an individual is zero that is the death distribution per age. *The task is to find the "mean health state" per age when we know the death distribution.*

Finding the Mean Health State as a summation of the stochastic processes of individuals could follow the lessons learned from the developments in Kinetic Theory in Physics. However, several new findings are needed. Especially as in the case of the human health, the data provided from the death distribution are produced from the health stochastic paths of the individuals when for the first time hit the zero health state barrier set at zero level or the X axis of a diagram.

That we search is first to find a distribution for the health state and then to estimate the final distribution when the health state paths hit for the first time the zero barrier. Then we can generate stochastic paths by stochastic simulations and verify the validity of our estimates. Both the direct and inverse problems are important tools to establish the Health State Theory.

**The Health State Defined, Modeled and Estimated**

That all people know by experience is that the health state is decreasing with age. By questioning we can have a scale for the health state ranging from 0 to 10 or as a percentage from 0 to 100. However, technically is simpler to accept the health state ranging from 0 to 1, with 1 been the "mean health state" of a population in the first years of the childhood and 0 at the age of the mean zero health of the population. This is the point where the number of deaths to the right is analogues to the number of deaths to the left of a graph.

An interesting comparison of the linear health state curve for the Mediterranean Flies studied by Weitz and Fraser and the curved with a negative slope for the human populations is also done along with stochastic simulations.

Torrance in the middle of 70s proposed a Health Status Index model suggesting the level of functioning of the health state of an individual at 1 for the perfect health state and lower values after injuries or diseases recovering after treatment to the perfect level in the first period of the life span. Then the level of functioning or health state is dropping down until the zero health level at the age of death (see Fig. 1).

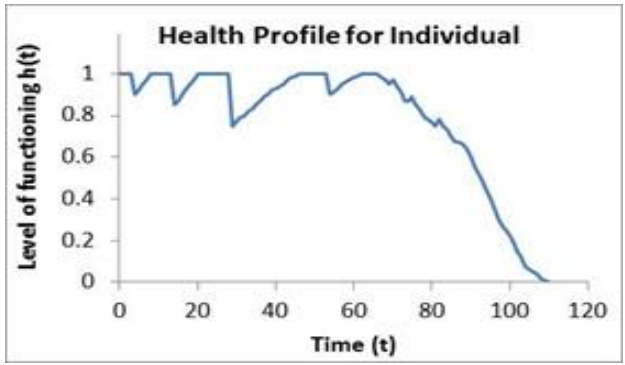

Fig.1. Based on: G. W. Torrance. "Health Status Index Models: A Unified Mathematical View", Management Sci., 22(9), 1976: 990-1001

Accordingly the health state of a population will be the average of a large ensemble of individuals, usually the total population of a country or a territory. Clearly this approach overcomes several shortcomings the main being the lack of a health state or health status unit of measurement. He accepts unity as the measure of the perfect health state for every individual. Clearly assuming unity as the maximum health state of an individual is not correct. Instead the "mean health state" should be expected to be at a unity level at maximum health state. The individual health state levels are expected to be at higher at lower values as it was the case of our modeling approach in 1995.

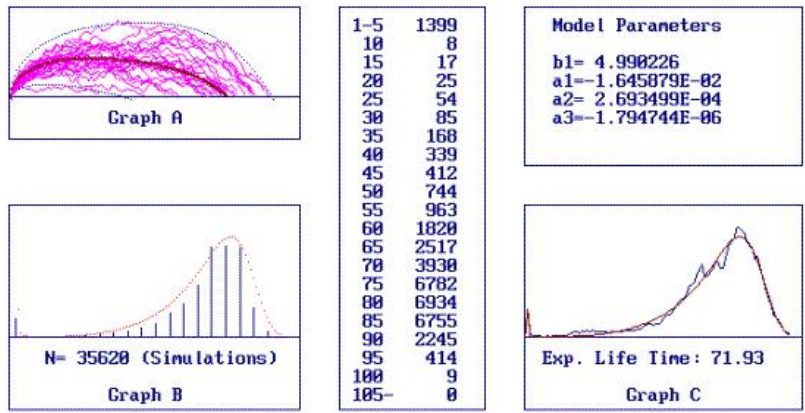

Fig.2. Based on: Janssen, Jacques and Skiadas, Christos, H. Dynamic modelling of life-table data, Applied Stochastic Models and Data Analysis, 11, 1, 35-49 (1995).

Fortunately, whereas the health state paths for the individuals is not possible to be estimated, the mean health state for the population can be found if we know the distribution of deaths that is the distribution that it is formed by counting the number of deaths at every age (usually estimated in yearly time periods). Then the first exit time theory of a stochastic process (expressing our health) crossing a barrier (here is the X axis representing the zero health level) provides the mean health state curve that is the health state of the population as a relatively smooth curve starting from a low level at birth reaching a maximum level and then gradually declining until zero (see Fig. 2). This was solved in 1995 by Janssen and Skiadas. In the same publication the inverse problem was approached that is to find the death probability density by generating a large number of stochastic paths by stochastic simulations. The general theory was developed in order to apply in the case of human mortality data.

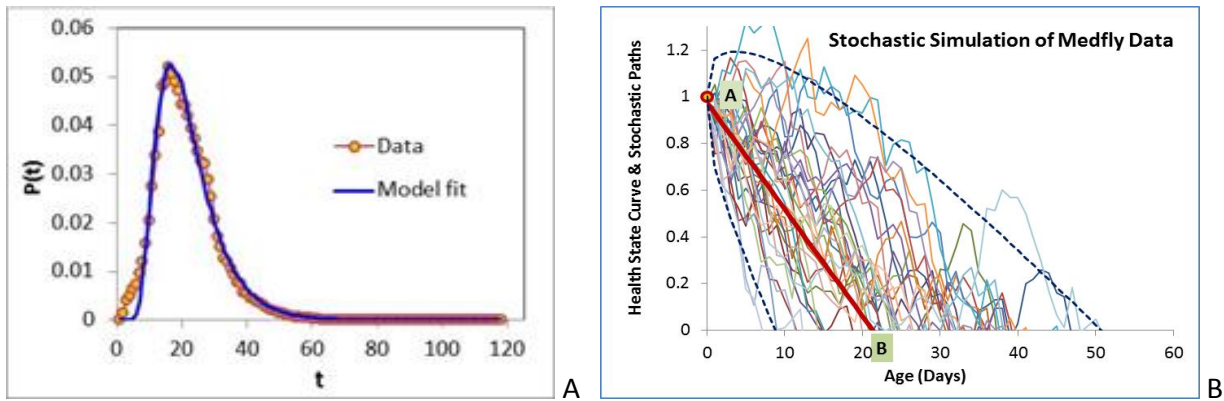

Fig. 3A. Based on: Weitz, J.S. and Fraser, H.B. Explaining mortality rate plateaus, Proc. Natl. Acad. Sci. USA, 98(26), 15383 (2001).

Fig. 3B. Our Simulations from the book "Skiadas, C.H. and Skiadas, C. Exploring the Health State of a Population by Dynamic Modeling Methods, Springer, 2017".

Few years later (2001) Weitz and Fraser applied the simpler first exit time model to the death data provided by Carey in a publication in Science. The health state curve for this case is a line starting from the level one and declining until the zero level (see Fig. 3A). We have done stochastic simulations (see Fig. 3B) to reproduce the data whereas we have estimated the parameters by fitting the model to data.

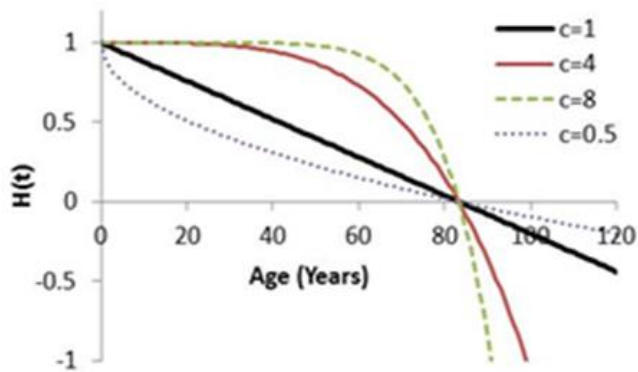

**Health State Function**

$$H(x) = 1 - (bx)^c$$

**Death Distribution**

$$g(x) = \frac{|H_x - xH'_x|}{\sigma\sqrt{2\pi x^3}} e^{-\frac{H_x^2}{2\sigma^2 x}}$$

$$g(x) = \frac{|1 + (c-1)(bx)^c|}{\sigma\sqrt{2\pi x^3}} e^{-\frac{\left(1-(bx)^c\right)^2}{2\sigma^2 x}}$$

Fig. 4A. Based on: Skiadas, C. and Skiadas, C.H. Development, Simulation and Application of First Exit Time Densities to Life Table Data, Communications in Statistics 39, 2010: 444-451.

Fig. 4B. Health State Model and Distribution of Deaths as a function of Health State $H(x)$ at age $x$. From Skiadas and Skiadas 2010, 2014, 2015, 2016, 2017.

The stochastic simulations for USA 2010 (females) are provided based on a model published in 2010. In this model the simple linear case of the Weitz and Fraser is expressed with $c=1$ whereas higher values for the exponent $c$ account for human mortality modeling (see Fig. 4A) to compensate for the repairing mechanisms of the human body. While the linear decline with $c=1$ is acceptable for medflies higher values for *c* are accepted for humans thus the mean health state tends to follow a rectangular like form well known in demography as rectagularization. The very simple equation form (see Fig. 4B) for the mean health state versus age *x* or operational state is: $H(x)=1-(bx)^c$. The related distribution function for deaths *g(x)* is included in Fig. 4B where σ stands for the standard deviation or the stochastic parameter. This parameter is important to reproduce the stochastic paths (see Fig. 3B and Fig. 5). Note that the derived distribution function is a new one perfectly fitting to the human mortality datasets. Technically, given the death distribution *g(x)* we fit the model to data by nonlinear regression to estimate the parameters *b*, *c* and σ.

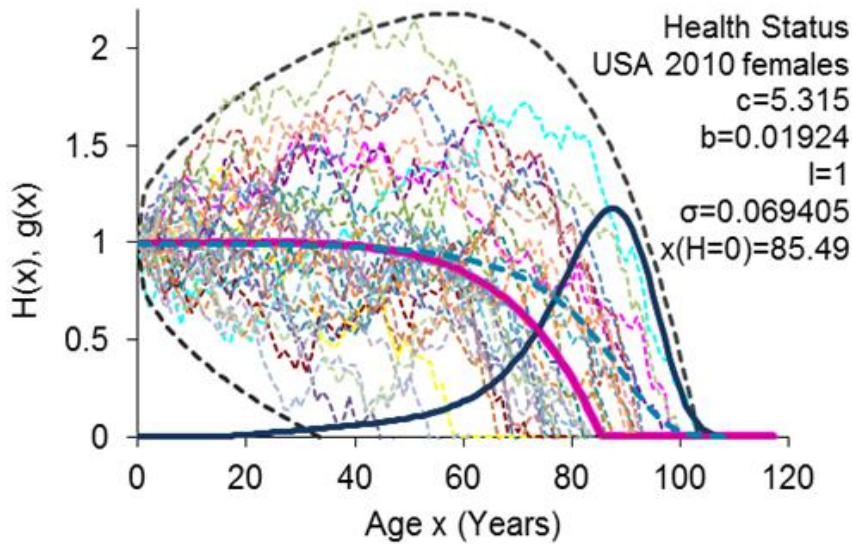

fig. 5. Our Simulations from the book "Skiadas, C.H. and Skiadas, C. Exploring the Health State of a Population by Dynamic Modeling Methods, Springer, 2017".

The main parts of the theory are illustrated in Figure 5A presenting how starting from the death distribution function (blue curve) we can estimate the Health State parameters.
The figure includes the Health State Function expressed by the heavy magenta curve, hitting the zero line at 85.49 years of age. This age is smaller to the modal age at death that is the age year with the maximum number of deaths. The heavy dark blue curve expresses the death density without the infant mortality cases.

Although the Survival Curve (cyan dashed curve) is known as long as the life tables have introduced, the Health State Curve was calculated after the introduction of the advanced stochastic theory of the first exit time.

The health state curve is illustrated by the heavy magenta line. The corresponding survival curve for the related case is presented by the cyan curve. The blue curve expresses the death distribution. The light curves with various colors are the stochastic paths from the related simulation. The two dashed black curves express the confidence intervals. Further to the Health State and the Life Expectancy, the age at mean zero health state is also estimated.

Fig. 6A presents an alternative method to reproduce the death probability density from stochastic paths and Fig. 6B provides the death probability density simulated (see Skiadas and Skiadas 2010).

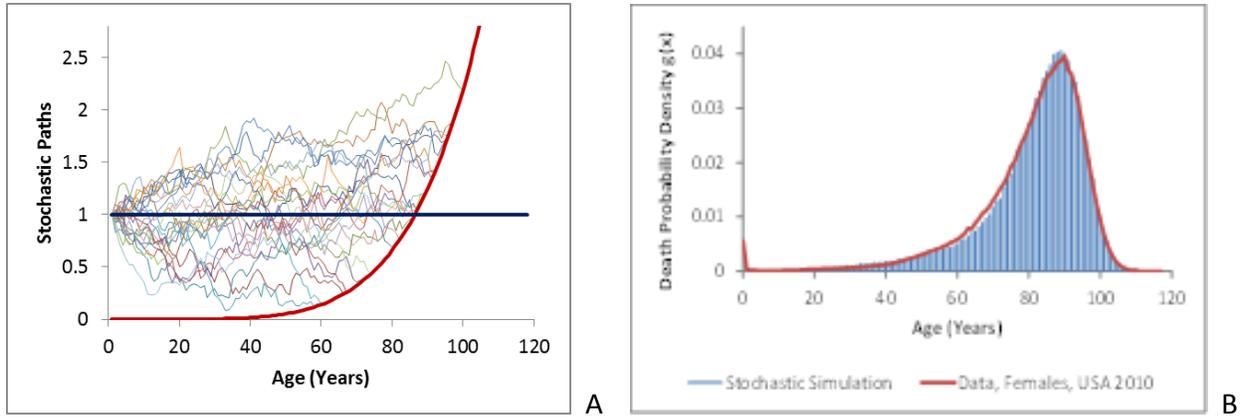

Fig. 6A. An alternative method to reproduce the death probability density from stochastic paths
Fig. 6B. Death probability density simulated (USA 2010, females)

As it was expected the introduction of a new tool as the health state function would lead to new advancement. The health state function estimated is a smooth declining function as is presented for males and females in USA (2010). We have estimated at which age the curvature is at maximum level. This is very important because this point is related with the maximum level of the human deterioration stage. In USA (2010) males this is achieved at 76.45 years of age and at 78.37 years for females (See Fig. 7A). The next graph illustrates the Health State Curve (red line), the Survival Curve (cyan line) and several survival cases for various values of the standard deviation $\sigma$. In the total rectangularization case the survival curve approaches the ABDC blue line (See Fig. 7B).

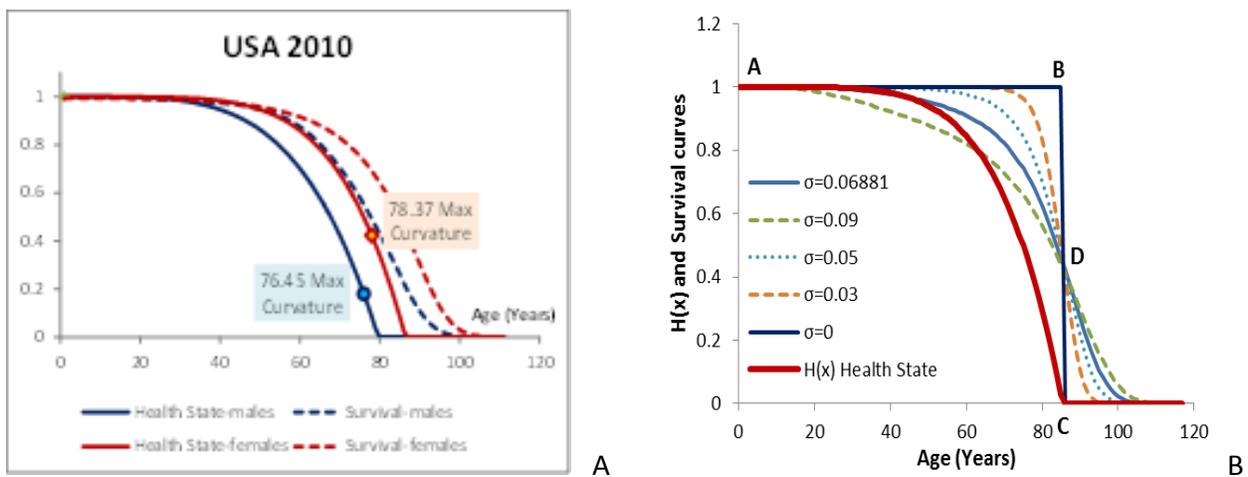

Fig. 7A. and 7B. Our estimations from the book "Skiadas, C.H. and Skiadas, C. Exploring the Health State of a Population by Dynamic Modeling Methods, Springer, 2017".

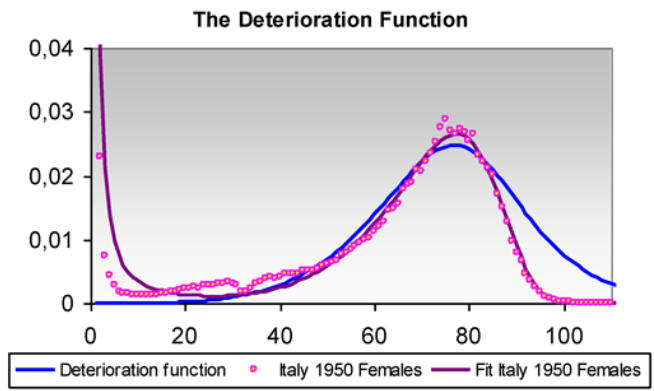

Fig. 8. Our estimations from the book "Skiadas, C.H. and Skiadas, C. Exploring the Health State of a Population by Dynamic Modeling Methods, Springer, 2017".

Note that the deterioration function has the form of a non-symmetric distribution function as in the above case for Italy (1950, females) rising slowly at the first ages and exponentially from the middle ages with a maximum at high ages close to eighty years and then declining at very high ages with an asymptotic decay explaining the Greenwood and Irwin (1939) argument for a late-life mortality deceleration or the appearance of mortality plateaus at higher ages (See Fig. 8).

Another important point is to find the full form of the human health state (see the figure for USA males the year 2000). The health state starts from a low level at birth grows to the maximum level one at 12 year of age declines to a local minimum at 22 years of age; then a local maximum is reached at age 32 and a continuous decline follows (See Fig. 9A). In Fig. 2B the two stage estimation is presented. The Health State Simple model H(x) is illustrated by the dashed orange curve whereas the blue curve represents the final form of the estimates.

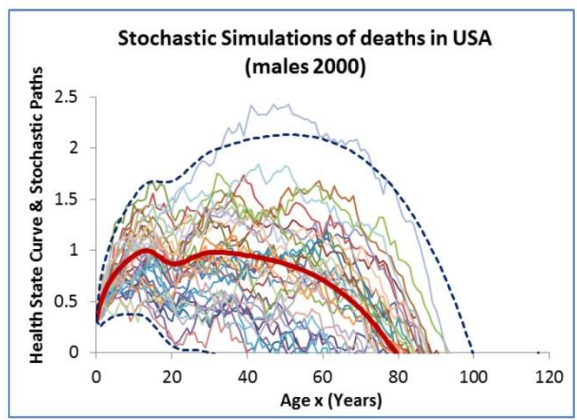 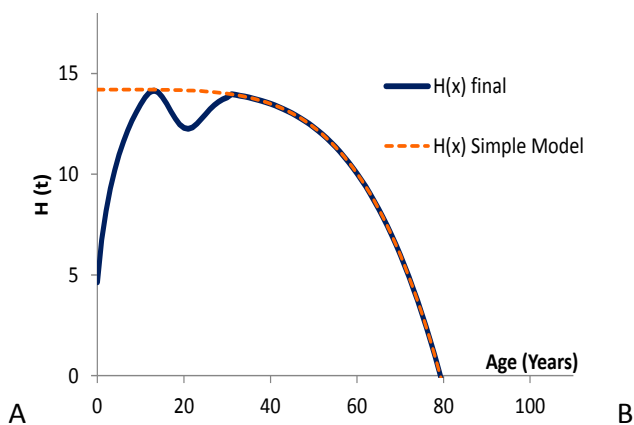

Fig. 9A and 9B. Our estimations from the book "Skiadas, C.H. and Skiadas, C. Exploring the Health State of a Population by Dynamic Modeling Methods, Springer, 2017".

The total health state is found by estimating the area under the health state curve (see Fig. 10A). The result is expressed as years of age that is 68.22 for males and 72.33 for females for USA 2010. The total health state (cyan curve) for females in Sweden (1970-2010) is presented (See Fig. 10B) along with our estimates for the healthy life expectancy (red curve) and the HALE estimates of the World Health Organization (rhombus with confidence intervals).

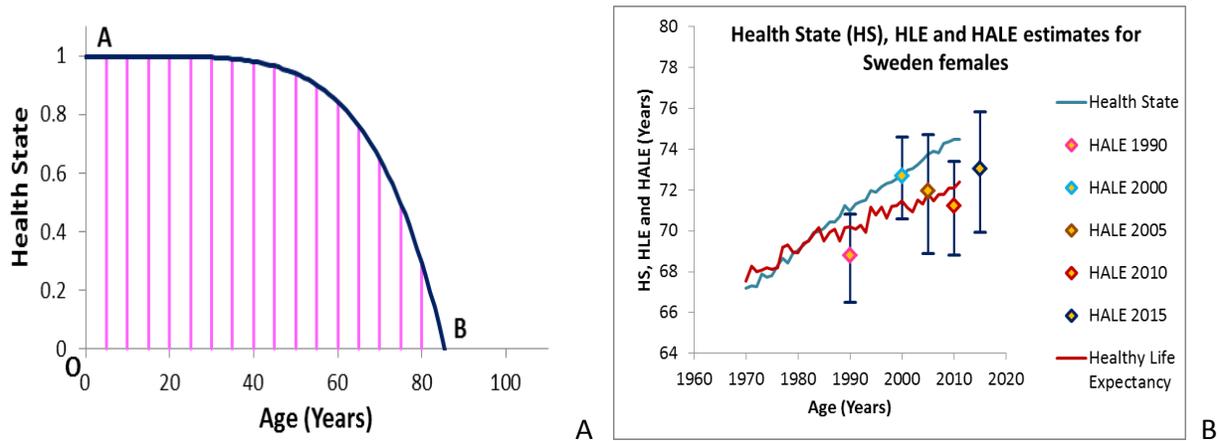

Fig. 10A and 10B. Our estimations from the book "Skiadas, C.H. and Skiadas, C. Exploring the Health State of a Population by Dynamic Modeling Methods, Springer, 2017".

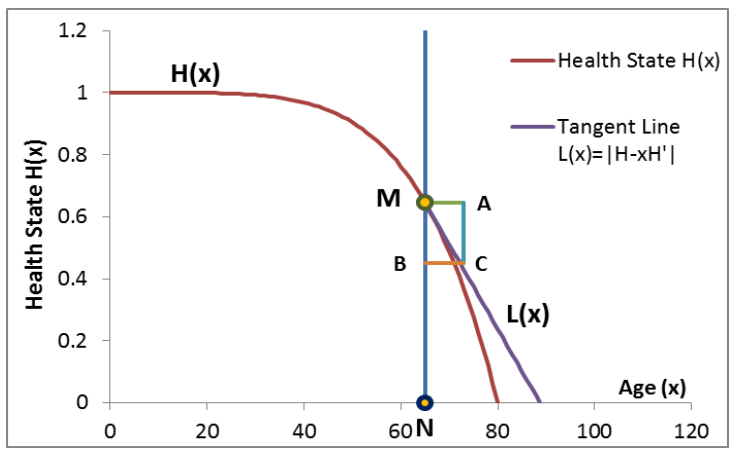

**Health State Function**

$$H(x) = 1 - (bx)^c$$

**Death Distribution**

$$g(x) = \frac{|L(x)|}{\sigma\sqrt{2\pi x^3}} e^{-\frac{H_x^2}{2\sigma^2 x}}$$

$$g(x) = \frac{|H_x - xH'_x|}{\sigma\sqrt{2\pi x^3}} e^{-\frac{H_x^2}{2\sigma^2 x}}$$

Fig. 11. Derivation of the Health State probability density function from the Inverse Gaussian

Fig. 4B. Health State Model and Distribution of Deaths as a function of Health State $H(x)$ at age $x$. From Skiadas and Skiadas 2010, 2014, 2015, 2016, 2017.

After introducing the Health State Function for the human population a very important point arises of finding a simple method to derive the probability density function based on the simple Inverse Gaussian presented earlier in the application of Weiss and Fraser for Medflies and already known from more than a century. A detailed methodology based on the stochastic theory is already presented in our publications mentioned above and in the references. The very simple transformation comes from the above figure 11A. We are dealing with health state *H(x)* at age *x* in point M where we approximate the curved part by the linear MC thus obtaining a tangent approximation. By moving the coordinate of the X axis to the point N at age *x*, the tangent line is given by *L(x)=|H-xH'|* . Then the death probability density function *g(x)* arises as a first approximation of a simple linearization of the Health State Curve at point M where the curve is replaced by the Tangent Line *L(x)=|H-xH'|* that is by a linear part of *H(x)* in the vicinity of the point M. Then the Inverse Gaussian applies for a small interval around the point M thus obtaining the function presented in figure 11B. This is the extended form we already have derived and applied for the health state of the human population. The related part will be included in a book in progress to appear in The Springer Series on Demographic Methods and Population Analysis.

**Further applications and results**

We have developed a method for estimating the Health State or the Viability of a Population in a time period from the distribution of deaths by applying the stochastic theory for the first exit time of a stochastic process. The theory developed is briefly presented in the Poster at the end of this paper (Figure 13) and included in our recently published book by Springer along with other interesting applications as the estimation of the healthy life years lost to disability and the maximum human life span. A part of the latter study is illustrated in Fig. 12 related to the female supercentenarian deaths, fit and forecasts in USA. The fit and projections for the 1980-2014 female deaths in USA approach the maximum year of female supercentenarian in USA at 119 years of age (IDL is the International Database on Longevity).

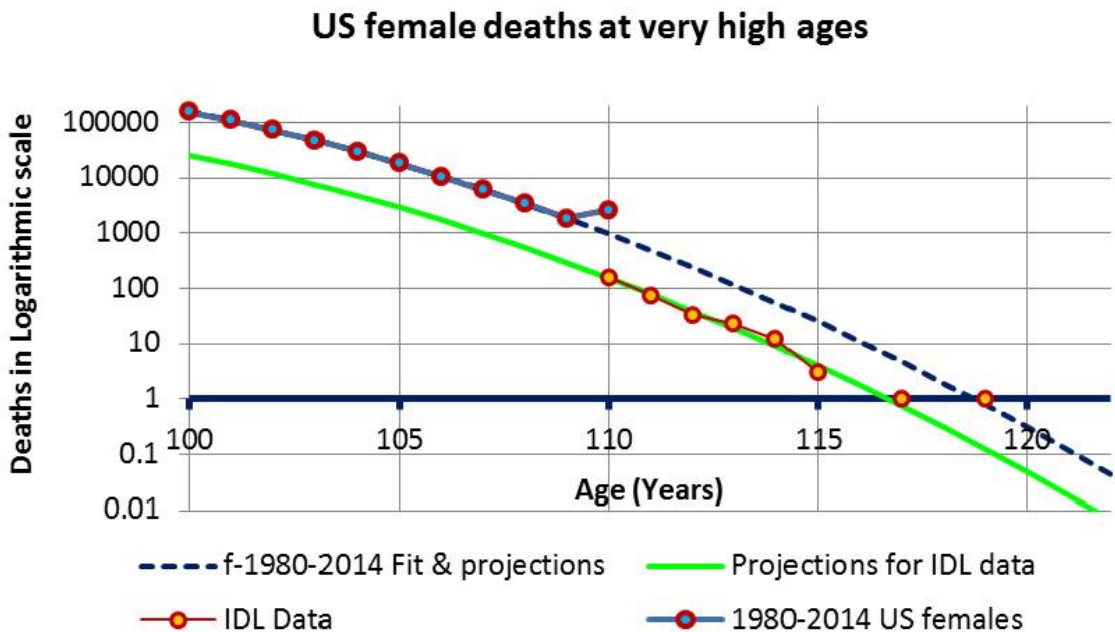

Fig. 12. Female supercentenarian deaths, fit and forecasts in USA.

Fig.13. Poster presenting the History of Health State Curves and related applications.